# A single-phase bcc high-entropy alloy in the refractory Zr-Nb-Ti-V-Hf system

M. Feuerbacher[a*], T. Lienig[a], C. Thomas[a]
[a] Ernst Ruska-Centre for Microscopy and Spectroscopy with Electrons and Peter Grünberg Institute, Forschungszentrum Jülich GmbH, 52425 Jülich, Germany

**Abstract**
We report on the production and characterization of a high-entropy alloy in the refractory Zr-Nb-Ti-V-Hf system. Equiatomic ingots were produced by arc and levitation melting, and were subsequently homogenized by high-temperature annealing. We obtained a coarse-grained, single-phase high-entropy alloy, with a homogeneous distribution of the constituting elements. The phase is a chemically disordered solid solution, based on a bcc lattice with a lattice parameter of 0.336(5) nm.

High-entropy alloys (HEAs) constitute a new field in materials science, dealing with alloys containing five or more metallic elements in equiatomic or near-equiatomic composition, which solidify as a solid solution on a simple crystal lattice [1, 2]. In these materials topological order and chemical disorder are present at the same time. This apparent discrepancy classifies them between conventional crystals, which possess chemical as well as topological order, and metallic glasses, which are disordered in all regards.

A prerequisite for the fundamental understanding of the consequences of these salient structural features for the physical properties, is the development of high-quality materials and their meaningful characterization. Experiments carried out on such well-defined samples can then readily be interpreted in terms of intrinsic materials properties, without misleading stray influences by secondary phases, undetected ordering, etc. To date, HEAs based on fcc, bcc and hcp crystal structures have been observed, for all of which equiatomic single-phase representatives exist, e.g. fcc FeCoCrMnNi [2], bcc ZrNbTiTaHf [3], and hcp HoDyYGdTb [4].

In the present paper we describe the production and characterization of a HEA in the system Zr-Nb-Ti-V-Hf. Our motivation is to explore an alternative bcc single-phase material, closely related to the well-known refractory HEA ZrNbTiTaHf, and make it available for comparative experiments. Two earlier publications relate to the Zr-Nb-Ti-V-Hf system, none of which however forestalls our results presented here. Li et al. calculated mechanical properties of a bcc ZrNbTiVHf HEA by *ab-initio* alloy theory [5]. In this purely theoretical work, the competing Laves phases were not taken into account, and the authors did not check whether or not a bcc ZrNbTiVHf phase forms in reality. Gao et al. [6] described the six-element system Zr-Nb-Ti-Ta-Hf-V but did not investigate or discuss the Zr-Nb-Ti-V-Hf subsystem.

On the long run, we strive toward the development of a single-crystal growth route for a bcc HEA. To date successful single-crystal growth was only reported for fcc HEAs [7] and bcc/B2 two-phase materials [8]. The known bcc HEA ZrNbTiTaHf has a very high melting temperature (2249 °C calculated as weighted average of the constituting elements [3]), which essentially limits the applicable growth techniques to crucible-free approaches. We have previously attempted to grow single crystals of ZrNbTiTaHf HEAs using the zone-melting technique, but could not effectively reduce the number of grains [9].

---
[*] Corresponding author.
*E-mail address*: m.feuerbacher@fz-juelich.de (M. Feuerbacher).

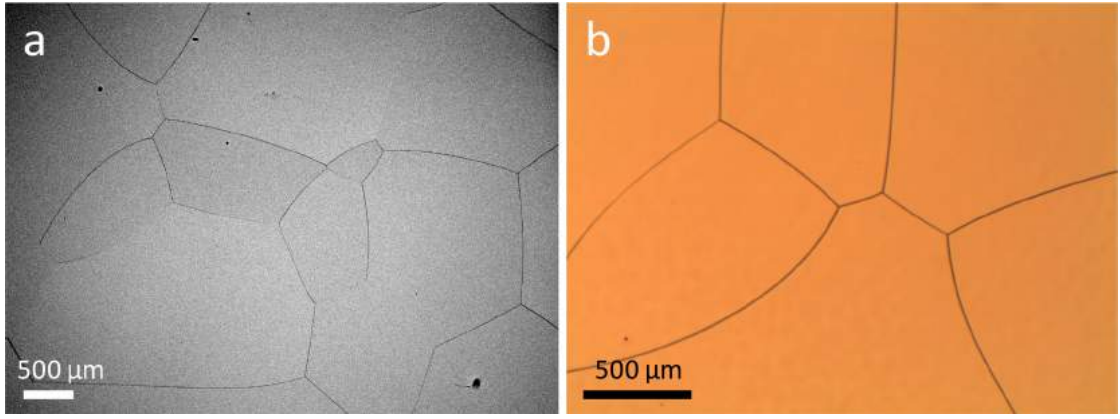

*Fig. 1: SEM micrograph using a backscattered-electron detector (a), and optical microcgraph (b) of the etched surface of a ZrNbTiVHf sample.*

The substitution of Ta by V reduces the melting point considerably (2025 °C calculated as weighted average of the constituting elements). The trade-off of this substitution is that with the inclusion of V, the competing C15 Laves phases $V_2Zr$ and $V_2Hf$ Laves phases have to be taken into account. Therefore, any growth of HEAs in the Zr-Nb-Ti-V-Hf system requires thorough characterization of the products.

The tendency of an alloy to form a HEA is commonly assessed considering two criteria: First, the total mixing enthalpy, calculated as [10]

$$\Delta H_{mix} = 4 \sum_{i<j} c_i c_i \Delta H_{ij} , \qquad (1)$$

should be small. Here $c_i$ and $c_j$ are the concentrations of the ith and jth element, respectively, and $\Delta H_{ij}$ is the binary mixing enthalpy of elements i and j. For a given alloy, the mixing enthalpy can be calculated using the tabulated $\Delta H_{ij}$ data provided e.g. in [11]. Values between – 15 and 5 kJ/mole are considered acceptable for the formation of a single-phase HEA [12].

Second, the radius difference $\delta r$ between the included atomic species should be small. According to [10]

$$\delta r = \sqrt{\sum_i c_i (1 - r_i/\bar{r})^2} , \qquad (2)$$

where $r_i$ are the radii of the individual atom species, and $\bar{r} = \sum_i c_i r_i$ is the weighted average radius. According to Wang [12], the radius difference for the formation of HEAs should be less than about 6.6 %.

For the case of an equiatomic alloy in the system Zr-Nb-Ti-V-Hf, we obtain a radius difference of 6.29%, and a mixing enthalpy of 0.16 kJ/mole. Hence, taking only these two formation criteria into account, the formation of a HEA in the system Zr-Nb-Ti-V-Hf can be expected.

However, the system has to be considered with more care, since in the subsystems V-Zr and V-Hf C15 Laves phases $V_2Zr$ and $V_2Hf$ exist, the formation of which competes with the formation of a pure solid solution.

In order to take competing Laves phases into account, we calculate Gibbs free energy of mixing

$$\Delta G_{mix} = \Delta H_{mix} - T\Delta S_{mix}, \qquad (3)$$

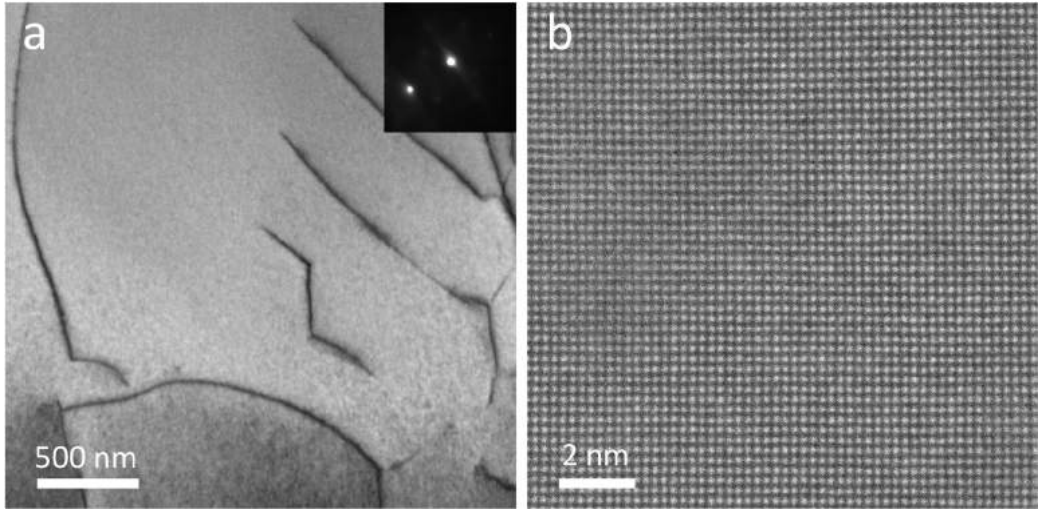

*Fig 2: TEM bright-field Bragg-contrast image (a), and HAADF-STEM image along the [1 0 0] lattice direction (b).*

where
$$\Delta S_{mix} = -R \sum_{i=1}^{n} c_i ln c_i \qquad (4)$$

is the entropy of mixing and $T$ the melting temperature of the alloy.

At the melting temperature, which was calculated as 2298 K by composition-weighted averaging the melting points of the constituting elements, and with the mixing entropy $\Delta S_{mix}$ = 13.38 J/mole for the equiatomic five-component alloy, we obtain $\Delta G_{mix}$ = -30.58 kJ/mole.

We compare this single-phase HEA formation scenario with that of the formation of two phases, a Laves phase and a solid solution of the remaining alloy. The mixing enthalpies of the possible $V_2Zr$ and $V_2Hf$ are -4 and -2 kJ/mole, respectively [11]. For a "worst-case" scenario, we consider the formation of $V_2Zr$, which has the lower mixing enthalpy of these phases. Assuming that for this ordered phase the mixing entropy is zero, we obtain a Gibbs free energy of mixing for the Laves phase of –3.56 kJ/mole using eq.(1). Assuming that all V will be used in the formation of Laves phase, the composition of the remaining solid solution is $Zr_{0.5}NbTiHf$. For this alloy, the mixing enthalpy according to eq.(1) is 2.62 kJ/mole, the mixing entropy is 11.24 J/mole K, and the melting point of the alloy is 2357 K. Using eq.(3), this combines to a Gibbs free energy of mixing of the solid solution of -23.88 kJ/mole. The total Gibbs free energy of mixing for the two-phase mixture thus amounts to -27.44 kJ/mole. This energy estimation hence suggests that the formation of a five-element solid solution is energetically favorable over the formation of a two-phase mixture of the Laves phase $V_2Zr$ and a solid solution of the residual elements by about 3.14 kJ/mole.

We have prepared equiatomic ingots using high-purity elements of Zr, Nb, Ti, V, and Hf by means of an arc-melting apparatus. The ingots were remolten three times using a high-frequency levitation melting apparatus in order to achieve a good mixing of the constituting elements. Subsequently the ingots were annealed at 1500°C for 6 hours in order to homogenize the material.

Samples for investigation by optical microscopy and scanning electron microscopy (SEM) were cut from the ingots by spark erosion. The surface was subsequently wet ground using paper up to 4000 grit, and polished using a Buehler MasterMet suspension on a soft cloth. For imaging the grain boundaries, the surface was additionally etched using a $HNO_3$/HF solution.

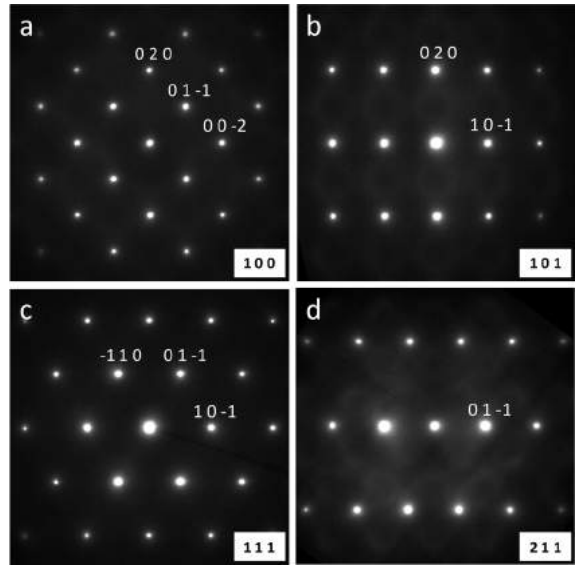

*Fig. 3 Partially indexed selected-area electron diffraction patterns along the [1 0 0] (a), [1 0 1] (b), [1 1 1] (c), and [2 1 1] (d) zone axes.*

SEM was carried out using a JEOL 840 instrument equipped with a backscattered-electron detector. These samples were also characterized using a Bruker D8 X-ray diffractometer (XRD) operated using Cu K$\alpha_1$ radiation. Specimens for transmission electron microscopy (TEM) and scanning transmission electron microscopy (STEM) were prepared using a FEI Helios Nanolab 400 S focused ion-beam system [13]. TEM investigations were carried out using a Philips CM20 microscope for analysis of the microstructure and for electron diffraction. For high-resolution imaging and EDX mapping a FEI Titan 80 – 200 equipped with an in-column Super-X EDX unit and high-angle annual dark-field (HAADF) detector [14] was employed. Chemical analysis was carried out on two samples of 180 mg taken from different areas of an ingot by inductively coupled plasma optical emission spectroscopy (ICPOES)

The composition as determined by ICPOES is $Zr_{20.8}Nb_{19.8}Ti_{19.8}V_{20.2}Hf_{19.3}$, which is very close to the nominal composition of the alloy. XRD diffraction of the sample shows a series of peaks that all can be indexed on the basis of a single bcc phase with a lattice parameter of 0.336(5) nm. This is in accordance (with a deviation of 6 %) with the theoretical lattice parameter of 0.3142 nm, calculated by averaging over the lattice parameters of the constituting elements.

Fig. 1 a depicts a micrograph of the etched surface of a ZrNbTiVHf sample taken in the SEM using a backscattered-electron detector. We find a single homogeneous phase with no indication of the presence of secondary phases. A few small grown-in pores and a network of grain boundaries can be seen. The average grain size is about 2 mm, with individual grains ranging from about 500 μm to 4 mm. Optical microscopy (Fig. 1 b) also shows the grain boundaries and the homogeneous interior of the grains.

The homogeneity of the material is further confirmed by TEM bright-field Bragg-contrast imaging (Fig. 2 a), which shows an even contrast and no secondary phases. Imaging under two-beam conditions reveals the presence of a moderate density of dislocations. The figure depicts a micrograph obtained using the (1 2 -1) reflection (inset) close to the [1 0 1] zone axis.

Fig. 3 displays selected-area electron diffraction patterns along the [1 0 0], [1 0 1], [1 1 1], and [2 1 1] zone axes. Again, the diffraction patterns can consistently be indexed on the basis of a bcc lattice.

The HAADF STEM micrograph displayed in Fig. 2 b was taken along the [1 0 0] direction. The image contrast corresponds to a regular square lattice. All bright spots, corresponding to atom columns, have approximately the same intensity, there is no long- or short-range ordering. In consistency with our previous results, this indicates the presence of a homogeneous disordered solid solution. The lattice parameter determined from the high-resolution micrographs (averaged over 18 measurements) amounts to 0.346±0.020 nm, which is consistent with the XRD result within the errors of the measurement.

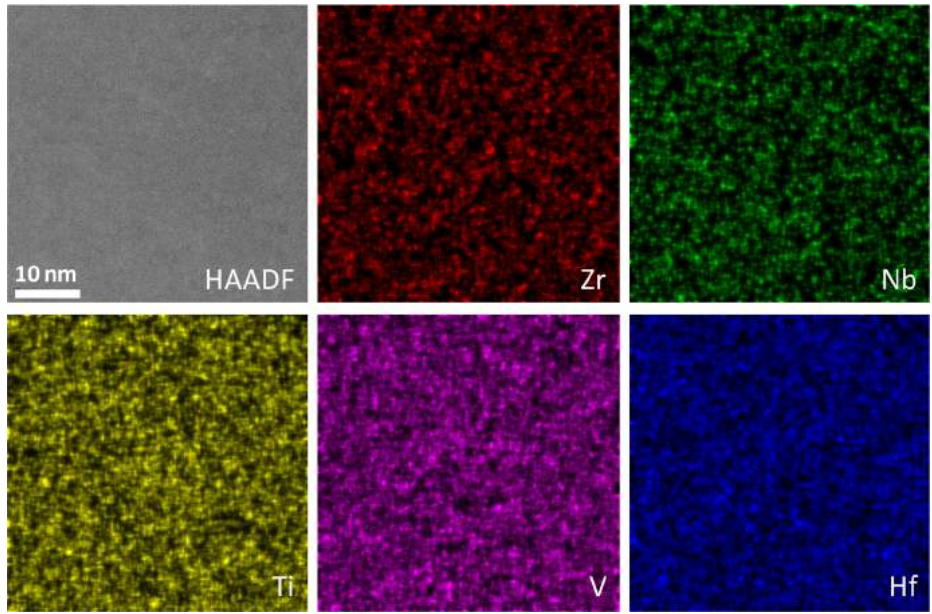

*Fig. 4: HAADF STEM image and EDX maps for all five constituting elements of the same sample area.*

Fig. 4 shows EDX maps for all five constituting elements along with a HAADF STEM image of the same sample area. All elements are evenly distributed, there is no indication of secondary-phase formation, ordering or clustering.

The melting point of equiatomic ZrNbTiVHf was found to exceed the high-temperature limit (1500 °C) of our DTA/DSC device. We therefore directly measured the melting temperature of a sample of the homogenized material, subjected to a constant power ramp in a high-frequency induction furnace, using an IRCON high-temperature ratio pyrometer. This revealed a melting temperature of 1590 ± 20 °C. This value is considerably lower than the calculated melting temperature of 2025 °C. In particular, this rather low melting point allows for a single-crystal growth approach by means of the Bridgman or Czochralski technique, due to the availability of suitable crucible materials in this temperature range.

In conclusion, we demonstrate that equiatomic ZrNbTiVHf is a homogeneous single-phase high-entropy alloy based on a bcc lattice. In agreement with our simple energetical considerations, a five-element solid solution phase solidifies, rather than a phase mixture including the competing C15 Laves phases. The described ZrNbTiVHf HEA represents an alternative to the well-known bcc HEA single phase ZrNbTiTaHf [3] and opens up the possibility for comparative measurements of physical properties of these two HEA materials. We believe that such experiments will lead to a deeper understanding of the intrinsic properties of single-phase HEAs and eventually allow conclusions on the structure-property relations.

The authors thank Dr. Marc Heggen for help with the STEM investigations, and Eva Würtz and Wilma Sybertz for discussions on sample characterization. This work was financially supported by the German Science Foundation (DFG) under Grant No. FE 571/4-1.